# The Bonds of Laughter:

# A Multidisciplinary Inquiry into the Information Processes of Human Laughter

**Pedro C. Marijuán***

**Jorge Navarro**

*Bioinformation and Systems Biology Group*

*Instituto Aragonés de Ciencias de la Salud. 50009 Zaragoza, Spain*

(*corresponding author) *pcmarijuan.iacs@aragon.es*

Abstract

A new core hypothesis on laughter is presented. It has been built by putting together ideas from several disciplines: neurodynamics, evolutionary neurobiology, paleoanthropology, social networks, and communication studies. The hypothesis contributes to ascertain the evolutionary origins of human laughter in connection with its cognitive emotional signaling functions. The new behavioral and neurodynamic tenets introduced about this unusual sound feature of our species justify the ubiquitous presence it has in social interactions and along the life cycle of the individual. Laughter, far from being a curious evolutionary relic or a rather trivial innate behavior, should be considered as a highly efficient tool for inter-individual problem solving and for maintenance of social bonds.

## 1. Introduction: the need of new synthetic views

The revival of laughter research during last two decades (Provine, 2000) has been very fertile concerning specialized achievements in the neuroimaging, neurophysiological, sound analysis, physiological (respiratory & phonatory), ethological, evolutionary, social and health aspects related to laughter. However, the conceptual counterpart of putting together the most relevant strands of thought in order to gain more advanced synthetic views or even to establish a new core hypothesis has not been developed sufficiently. This paper will attempt that –though, inevitably, in too idiosyncratic a way.

The preliminary idea is to establish a coherent link between the evolutionary roots of laughter and the origins of language, aligned with the "social brain" hypothesis (Allman, 1999, Dunbar, 2004). A perfunctory examination of laughter in the social communication context and along the life cycle of the individual will allow a first approach to the new argument–core hypothesis, and will present laughter as a "virtual grooming" and bond-making instrument. Afterwards, a behavioral correspondence with the underlying neurodynamic events (Collins & Marijuán, 1997) and a *sentic* hypothesis on the informational/emotional content of the different forms of laughter (Clynes, 1979) will be tentatively framed. Laughter, will be concluded, has been evolutionarily kept and *augmented* as an optimized tool for unconscious cognitive-emotional problem solving, and at the same time as a way to preserve the essential fabric of social bonds in close-knit groups and within human societies at large.

## 2. The evolutionary scenario of human laughter

Classical and recent ethological studies have unambiguously situated laughter within signaling contexts of play and socialization of "advanced" mammals, especially in relation with the grooming practices of anthropoid primates, but also in rodents and other species (Panksepp, 2005). Whether anthropoid ritualized "panting" during play should be considered as the closest antecedent of human laughter is still a matter of debate, factually settled down (Ross et al., 2009). Anthropoids (chimps) "laugh" mostly when tickled and at chased games, noisily punctuating each inhalation and exhalation; but they are fundamentally unable to modulate a single exhalation and articulate it into discrete notes. Human modifications upon this primate precursor of laughter have undoubtedly derived from the systemic adaptations involved in bipedestation, allowing an improved control of breathing by freeing the thorax of the mechanical demands of quadrupedal locomotion –and also freeing the hand with the subsequent emergence of human dexterity techniques, directly fuelling the neocortex expansion too. "In the beginning was the breath" (Provine, 2000).

New social behaviors were driving further evolutionary changes (mostly brain-centered) of the human species, and they presumably included an increase of group size and the development of articulate communicative language, with decoupling of vocal production from emotions. New feeding practices and an improved social sharing of food (including the crucial invention of exodigestion or "cooking") were also needed to compensate for the "energy crisis" that so large a brain was causing in the metabolic budget, probably already at the level of *Homo ergaster* (Allman, 1999; Wrangham, 2009). Actually, an evolutionary trade-off took place between gut tissues and brain tissues: the great expansion of the brain in humans was accompanied by a commensurate reduction in digestive organ weight, almost "gram-by-gram" (Allman, 1999).

The loss of bodily hair was behaviorally important too, both for heat dissipation in new hunting strategies based on long-distance running needed for the new diet, and for the appearance of new pair-mating behaviors and a stronger parental bonding (Jablonski, 2010); it further facilitated the evolution of new sexual signals, which were also accompanied by many other group communicational adaptations: laughter, crying, facial expressions, blush, pallor, enhanced gaze discrimination, unison sense, rhythm, music, dance... (Benzon, 2001).

## 3. The Social Brain hypothesis

In the above evolutionary overlapping of highly consequential positive feedbacks, both of physical and behavioral nature, a crucial correlation occurs between social life and brain development. Concretely, among the different primate societies, one of the most significant evolutionary correlations appears between the group size and the relative neocortex size (Dunbar, 1998). See Figure 1. The idea of relating brain size with the demands of communication in social life, already hinted by Darwin, was framed as a social hypothesis in the 80's and early 90's by Allman and others; it was also dubbed as the *Machiavellian intelligence hypothesis* by Byrne and Whiten (see Allman, 1999). Later on it was more rigorously formulated by Dunbar and extended into other mental fields by Baron-Cohen, Badcok and Crespi (see Dunbar 2004; Badcok & Crespi, 2008).

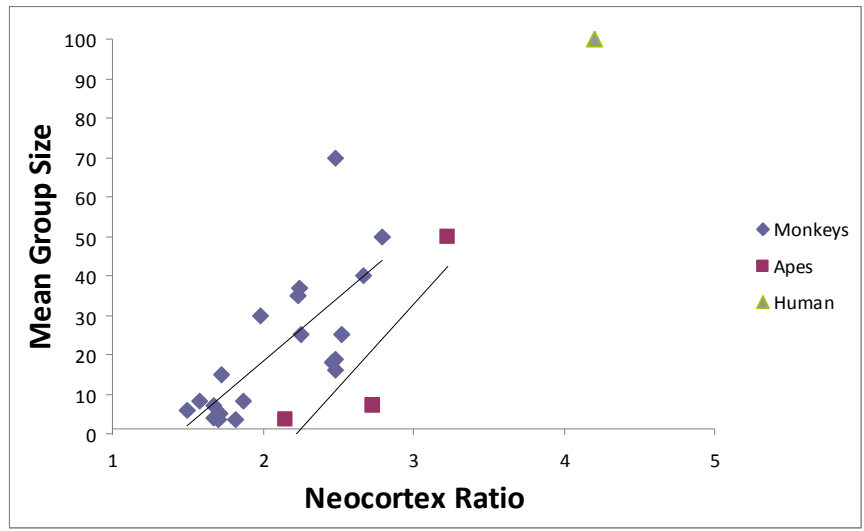

**Figure 1.** Representation of the mean social-group size in monkeys and apes (ordinates) versus the relative neocortex volume (abscises); in human species both data are disproportionably high. In the figure, diamonds represent monkeys, squares represent apes, and the triangle represents humans. Modified from Dunbar & Shultz (2007).

The social brain hypothesis posits that, in primate societies, selection has favored larger brains and more complex cognitive capabilities as a mean to cope with the challenges of social life (Silk, 2007). Contrary to conventional wisdom in the cognitive field and neuroscience, which assumes that animal and primate brains deal with basically ecological problem-solving tasks, what the large primate brains would accumulate in their expanded neocortex is not information about ecological happenstances but the computational demands of their complicated storylines: the important memory capabilities invested in other individuals, the ever changing coalitions, the mating alliances, the sharing of resources, the multiple conflicts, and so on. Social networks in primates seem to be very different from those found in most other mammals: they are cognitive, memory-loaded, based on bonded relationships of a kind found only in pairbonds of other taxa (Dunbar & Shultz, 2007).

Maintaining that special structure of social-cognitive bonds relies on grooming practices. "Bonds" are but shared memories: they consist of neural *engrams* encoding behavioral interactions that have been finalized positively (Collins & Marijuán, 1997). When altered in the behavioral "noise" of primate societies, bonds are rebuilt and emotionally restored throughout a variety of grooming practices: touching, scratching, tickling, playing, massaging… up to 20 % of ecological time may be devoted to participation in grooming networks. The molecular cocktail activated by grooming is intriguing and not quite solved yet. Seemingly, it involves neuropeptides and relaxing hormones of the neural reward system, with effects in stress quenching, immune boosting, and also in learning processes (Shutt et al., 2007; Nelson, 2007). These powerful neurotrophic mechanisms, very similar to those already authenticated for mammalian pairbonding (e.g., oxytocine, AVP –see Allman, 1999), would reinforce the involved synaptic memories and would restore the bonding relationships.

Frequent pair-wise grooming in between individuals, however, imposes a strict time limitation regarding group size: depending on diet, 20% of time is the upper ecological limit that grooming can reach. This factor necessarily restricts the size of grooming networks and, thus, of natural groups in primate societies (composed, at most, of a few dozen individuals). So, how could human societies have organized their "grooming" within increasingly larger natural groups, of around 100 or 150 individuals? As Dunbar (1998, 2004) has argued, human language was the evolutionary solution.

"Languaging" was co-opted as a *virtual system* for social *grooming-massaging*, plus other specifically human adaptations for group cohesion: laughter, crying, gaze-facial expressions, music, dance... It is by following this line of thought, that the enigmatic presence of laughter along the human life cycle may be further clarified.

## 4. Languaging and laughing

As a means of communication, language purports an instinctive simplicity that obfuscates the individual perception of its limitations. In the neurodynamics of lenguage, for instance, a fundamental transition occurs between talking and listening (Collins & Marijuán, 1997), between being "groomer" and being "groomed" in general, there is a slight behavioral preference for being the groomer. In each case, the reconfiguration of the involved neural systems proceeds along a very different branch regarding functional closure of the action/perception cycle.

The control of this neurodynamic talking/listening transition is at the same time a fundamental socio-cultural matter, heavily regulated depending on context, identity of the speakers, age, gender, hierarchy, etc. It is the complexity of "taking turns": half-conscious signals for the informal settings (families, friends, clubs, restaurants), and strict procedures in the formal settings (committees, seminars, conferences, ceremonies).

In the dynamics of a group conversation, frequent changes and instabilities occur as successive parties are added: 2, 3, 4, 5… Almost inevitably (unless formal or informal rules intervene) the ongoing conversation will split into smaller "partitions", very frequently of 2, 3, and 4 individuals. Statistically, the average talking group is of 3-4 individuals, within a maximum 10-12 of preferred clique size. The daily budget of conversation amounts to an average of 3-4 h, being "gossiping" and "small talk" preferred contents rather than the exchange of factual information (only 1/3 of time). These are very robust data, consistent in a variety of social and cultural contexts (Dunbar, 1998, 2004).

According to Dunbar's version of the social brain hypothesis, the previous conversational data dovetail with the grooming needs of human natural groups, around 3-4 times of bigger size than other anthropoid societies. Considering small talk as the social grooming of our species, it would provide thrice as much virtual grooming on average than the strictly bilateral physical grooming characteristic of primates. By means of the talking/listening exercise, individuals would impart each other a mental *massage*: amusing themselves, actualizing their relationships, gossiping about absent third-parties... in the long run maintaining the mutual bond. Human social networks so glued by the linguistic nexus will manifest a complex mixture of links: parenthood-related "strong bonds" and many other classes of more labile "weak bonds"–curiously, as happens in the biomolecular realm, weak links turn out to be the genuine bonds of social complexity, those in which the growth of civility is supported (Ikegami, 2005).

*Laughter in conversation*

Laughter quite often breaks in amidst the talking/listening exercise. Having evolutionarily preceded language, laughter has continued to fulfill very especial tasks regarding the communicational grooming of human groups. What has been called "antiphonal laughter" (the chorus of laughing people –see Smoski & Bachoroswki, 2003; Smoski, 2004) may be seen as an *effective extension* of the talking massage effects in bigger groups, where the mere size precludes active participation of most individuals in the talk; the laughing together that ensues, brings the augmented neuromolecular grooming-effects of laughter available to everybody in the group irrespective of the conversation share.

Laughter is regularly situated at the very end of verbal utterances; it punctuates sentences as a sort of emotional valuation or as an enigmatic social "call", even in deaf people using the hand-sign language (Provine & Emmorey, 2006). In this sense, laughter production, far from interfering with language or competing as a "low level" process with the higher cognitive functions for access to the fonatory apparatus, becomes itself a *cognitive solution*, marking the occurrence of humorous incongruences as positively finalized items within the ongoing talking/listening exchange.

During conversation exchanges between genders, laughter enters as a *bona fide indicator* gauging the relative advancement of bonding processes in courtship (males usually are providers of laughter, "groomers", while females are consumers, "groomees" –see Provine, 2000); laughter contributes as well as a lively tool in the establishment of parentofilial bonds (the babbling-laughing charms that babies and toddlers address to their parents). In general, the occurrence of laughter indicates that successful bonding processes of whatever type are in progress between the laughing individuals; it is the case of laughter addressed "against" someone outside the laughing chorus too. It is also the case of the evolutionary relationship between laughter and the "sharing of food". Presumably, the pleasurable grooming activities of languaging & laughing did coevolve as social bonding tools with the pleasurable "sharing of food" brought about along the exodigestion cultural practices of cooking. In every culture, eating together maintains a ritual significance as a bond-building occasion, usually full of small talk and antiphonal laughter episodes. (From this angle, contemporary "restaurants" are indeed feeding places, but even more they are group bonding places; the restaurant table in particular becomes a terrific scenario to follow the partitional dynamics of group conversation-transitions!)

## 5. The abstract neurodynamic "stuff" of laughter

The great variety of stimuli and situations conducing to laughter –physical, chemical, sensorimotor, cognitive, relational, parental, courtship, play, pathological, etc– and even more the intriguing neuromolecular repercussions of this innate behavior, are a warning of the sheer complexity of neurodynamic events underlying it. After almost two decades of neuroimaging works, for instance, almost any brain area has been related to laughter and humor; and like in the deepest cognizing problems, no decisive results have been found yet regarding a unified explanation.

*Neural pathways and systems*

Medically, the study of pathological laughter (Poeck, 1985) has pioneered the field respect other behavioral and cognitive approaches to "normal" laughter. Lesion studies (e.g., damage to frontal cortex areas) have pinpointed the participation of many specific areas in humor perception and laughter production, and have also dispelled much too simple an assumption. It has been authenticated that, unlike in emotional responses, relatively confined to specific areas, laughter is associated with activation of numerous regions: left, front, right, and rear of the cortex, as well as the motor areas, cerebellum, limbic system and subcortical nuclei, hypothalamus, etc. The classical view is that two main neural pathways, relatively independent, are controlling the expression of laughter (Ozawa, 2000; Iwase, 2002; Wild et al., 2003). The former is more "involuntary and emotional", and involves amygdalar, thalamo-hypothalamic, subthalamic, and dorsal mesencephalon areas; while the latter, more "voluntary and cognitive", originates in premotor/opercular frontal cortex, and links with the pyramidal tract and brain stem (Goel & Dolan, 2001). As Parvizi et al. (2001) have noted, a more comprehensive scheme can be elaborated that includes the loops associated to the cerebellum and responds better to the cases of pathological laughter. See Figure 2. Besides, it is interesting that systematic gender differences have been found regarding patterns of activation in cortical, hemispheric, and mesolimbic structures in response to humoristic stimuli (Azim et al., 2005); and that the mesolimbic structures activated by laughter and humor include the nucleus accumbens, a key component of the mesolimbic dopaminergic *reward system* (Mobbs et al., 2003). Clasical EEG studies have also generated an ample literature on cortical "wave" events accompanying laughter onset and perception of humorous stimuli and (Derks et al., 1997).

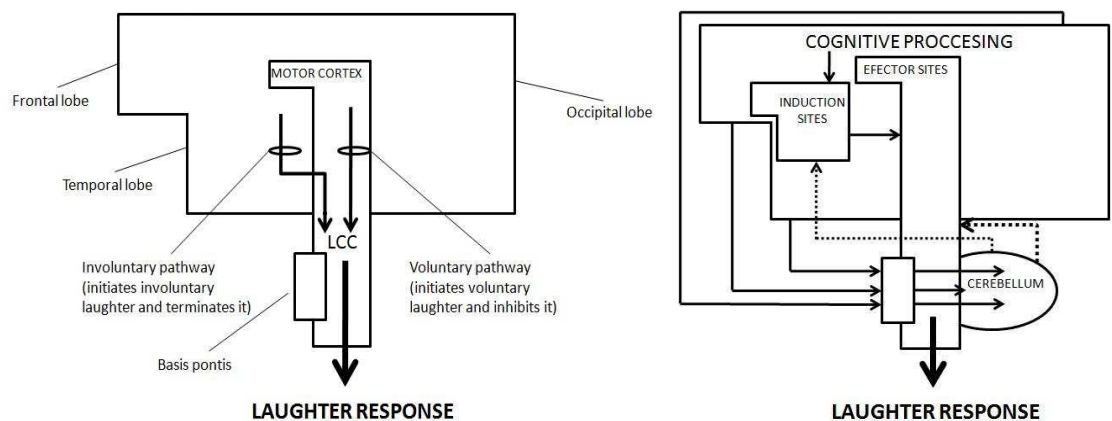

**Figure 2.** Left: the traditional view of laughter and crying circuits, emphasizing how two separate pathways conduce to activation of the specific laughter and crying center (LCC). Right: the cerebellum's role is emphasized, appearing as the processing center where a certain profile and level of emotional response is computed according to signals received from the telencephalic structures, in which emotional –competent stimuli as well as the relevant cognitive/social context are processed. Modified from Parvizi et al. (2001).

Certainly, a unified neurodynamic explanation should integrate the multitude of potentially participating areas and nuclei into functional constructs with behavioral sense. Catchword terms such as "species call",

"false alarm", "polarity change", "pathways collision", "release of tension", "collapse of strained expectation", and so on, have been historically proposed by scientists and philosophers to explain the role of laughter either in social, behavioral or neurodynamic grounds (Ramachandran, 1998; Provine, 2000). The point of view advocated here, in the nearness of S. Freud, A. Koestler, O. Rossler, and others (see Marijuán, 1999), attempts the exploration of the "minimization of incoherent excitation" motto for the explanation of laughter. It means relying on the conceptual track around the minimization (optimization) of structural and functional features of the vertebrate Central Nervous System (CNS). A substantial body of neuroscientific literature, starting with neurophysiologist Ramón y Cajal "Laws of Economy" (1899), has been developed during recent decades, including new ideas on optimality in the dynamics of connectivity among neural assemblies see Marijuán (2001), Edelman and Tononi (2000).

*Laughter and optimality*

In an extremely succinct way, the brain organizes its information processes by optimizing its own excitatory state throughout a global/local entropic variable (Turvey, 2004). Behavioral "problems" become destabilizing occurrences that are coded as patterns of uncorrelated excitation & inhibition gradients projected upon neuronal areas and nuclei –within the whole interconnected mappings arranged as a topological homeomorphism. By means of "active" minimization structures (cerebellum, thalamus, hippocampus, etc.) different rhythms and inhibitory flows are driven out upon the excitatory "problems", along successive reenactments of the action-perception cycle, until "solutions" are finally sculpted as entropy-minimized constellations, recorded then in the columns of the cortical memory banks. Problem solving, whatever its "level" (perceptual-motor, categorization upon cortical memories, advanced cognition), is accompanied by learning and by synaptic reinforcement, as well as by activation of the reward mechanisms (Collins, 1991; Collins & Marijuán, 1997).

Laughter becomes a quasi-universal information processing "finalizer". We laugh "abstractly": when a significant neurodynamic gradient vanishes swiftly, i.e., when a relatively important problem of whatever type has been suddenly channeled in a positive way, and has vanished as such problem. Like in the slow tension growth and fast release of physical massage, the paradoxical, or tense, or contradictory situation suddenly becomes a well-known case of pleasurable, primary, childish, stumbling, babbling, or retarded-foreigner nature. Problem solved! The "idle" excitation still circulating in the regular problem-solving of cortical and limbic structures is redirected towards the fonatory apparatus where it produces an unmistakable signature. It is the "call" of the species, a social signal of wellness after successful problem solving, after effective mental massage. The sound form of laughter would bear a trace on the kind of neurodynamic gradient that originated it (Marijuán, 2009).

In the extent to which this scheme is acceptable, or at least permissible as a heuristic approach, it can throw light on why humans have evolutionarily augmented the innate behavior of laughter (as well as crying and other group emotional adaptations). Laughter is spontaneously produced to minimize occurring problems in an automatic-unconscious way that mobilizes powerful neurodynamic and neuromolecular resources without any extra computational burden on the ongoing conscious processes of the individual. In the complex social world that the enlarged human brain confronts, with multitude of perceptual, sensorimotor, and relational problems, and above all with those derived from the conceptual-symbolic world of language in the making and breaking of social bonds, informational problems dramatically accumulate in very short time spans. Thus, it makes a lot of evolutionary sense counting with these extra-ordinary minimization resources: the information processing power of a hearty laugh (or of bursting out into tears!).

## 6. The sounds of laughter: revisiting the *sentic forms* hypothesis

Laughter and infant crying are two of the more potent, affect-inducing vocal signals (Bachorowski & Owren, 2005); they are "evolutionarily designed" as species-specific relevant auditory stimuli that immediately provoke emotion-related responses in any listener. (*En passant,* there is an intriguing symmetry between laughter and crying sounds, and also between their affective responses: Perhaps because they respectively imply the *making* versus the *breaking* of social bonds, the beginning of lasting memories versus the loss of important memory constructs?). Still unclear, however, where the auditory emotional clutch localizes inside these innate human sounds.

*Sound structures of laughter*

Far from being a stereotyped signal, laughter becomes one of the most variable acoustic expressions of humans, comparable to language except for the combinatory richness of the latter. Typical laughter is composed of separate elements or "calls" or "syllables", *plosives*, over which a vibrato of some fundamental frequency *Fo* is superimposed (Rothganger et al. 1997); a typical laughter episode may last around one second (or slightly less) and will contain around 5 plosives (most often, in between 2 and 8). An important distinction to make is between "vocalized" and "unvocalized" laughter; even though the former induces significantly more emotional responses in listeners, the latter appears consistently in many laughter records, comprising a large variety of sounds (snort-like, grunt-like, giggles, chuckles, etc.)

In a landmark experimental study, Bachorowski et al. (2001) found that there are around 4.4 calls or plosives within each laughter bout, a single plosive having a duration of 0.11 s and a separating interval of 0.12 s (for voiced laughter). Call or plosive production is denser towards the beginning of laugh bouts, and inter-plosive durations gradually increase over the course of bouts. The average value of the fundamental frequency *Fo* for male laughs is 272 Hz (sd = 148) while for females is considerably higher and more variable 405 Hz (sd 193); only for voiced laughs, the respective values are 282 and 421 Hz. Usually *Fo* is much higher in laughter than in speech, thus, extremes of male *Fo* were found to be as high as 898 and as low as 43 Hz, while female extremes were in between 2083 and 70 Hz. The excursions of *Fo* along the bout trajectory represent an additional factor of variability, showing contours such as "flat", "rising", "falling", "arched", sinusoidal", etc.

All of the previous elements could form part of the inbuilt cues to *laugher identity*, which have been proposed to play an important role in listener emotional responses (Baworowski & Owren, 2001). In particular, the pitch or *tone* curve described by *Fo*, together with the distribution of plosives, would show consistent differences between laugh forms associated with emotional states of positive and negative valence (Devillers & Vidrascu, 2007). The main trend is that the energy and duration becomes higher for "positive" than for "negative" laugh, and vice versa for the relative presence of unvoiced frames, more frequent in ironic and hostile laughs than in joyful ones. Notwithstanding that, there is not much consensus established yet –neither significant hypothesis to put to test– on how the interrelationship between plosives, tones, melodies and other variables of laughter may be systematically involved in encoding and distinguishing the underlying emotional states (Bea & Marijuán, 2003; Bachorowski & Owren, 2008).

*Connecting with the "sentic forms" hypothesis*

At this point, the *sentic forms* hypothesis, framed by M. Clynes in the 70's, could help in the exploration of new directions for such open questions. If laughter contains inner "melodies" or pitch patterns of emotional character, how could they be structured?

Following the sentic paradigm developed around tactile emotional communication by means of exchange of pressure gradients, there appears a set of universal dynamic forms that faithfully express the emotional interactions of the subjects (Clynes, 1979). The universality of these behavioral performances stems out from a common quality, a unique dynamic *essentic form* (or sentic, for short) that conveys the essential interactive information of each emotion. Moreover, irrespective of the sensory modality involved, or of the type of motor expression used, such patterns show a remarkable consistency. The nervous system is built in such a coherent way that it not only executes this dynamic form but also perceives it accurately and precisely. Subsequently, the whole set of sentic forms can be determined experimentally, and be measured, catalogued, etc. by means of the tactile expression of emotions; sentic forms can also be found reliably in musical phrases, facial expressions, and in the visual arts (Clynes, 1988, 1992). See Figure 3.

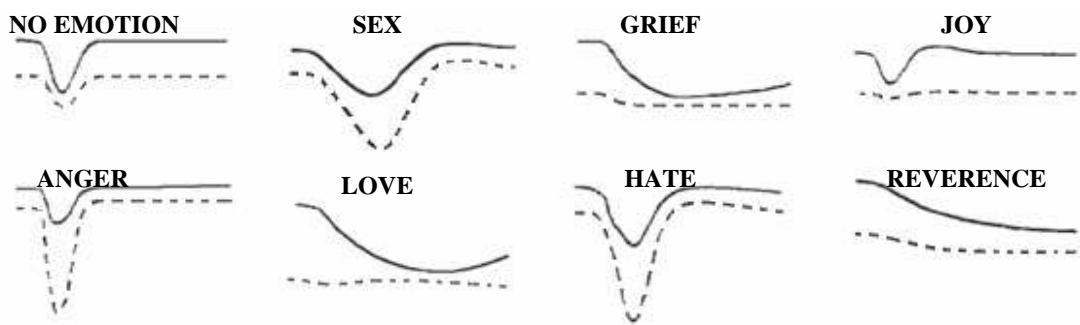

**Figure 3.** Representation of sentic forms. Each of these figures represents a differentiated emotion pattern of finger pressure obtained in laboratory from subjects who were asked to push a button in response to elicitations of eight different emotions. Figures are representing Pressure (0-200g/m$^2$) vs. Time (0-2s). The upper lines represent

downward-upward pressure, whereas the lower dashed lines represent forward-backward pressure. Modified from Clynes (1979).

Regarding the tentative application of sentic forms to laughter, two methodological changes are needed. First, an inverted representation of the sentic curves (so that positive increases of pressure read upwards), and second the introduction of some more precise mathematical formulations. Following the formal reasoning of D. Winter (1999), based on wave interference grounds, a series of mathematical expressions would characterize the four most important emotional-sentic expressions; among them the very *golden mean* would show up. See Figure 4.

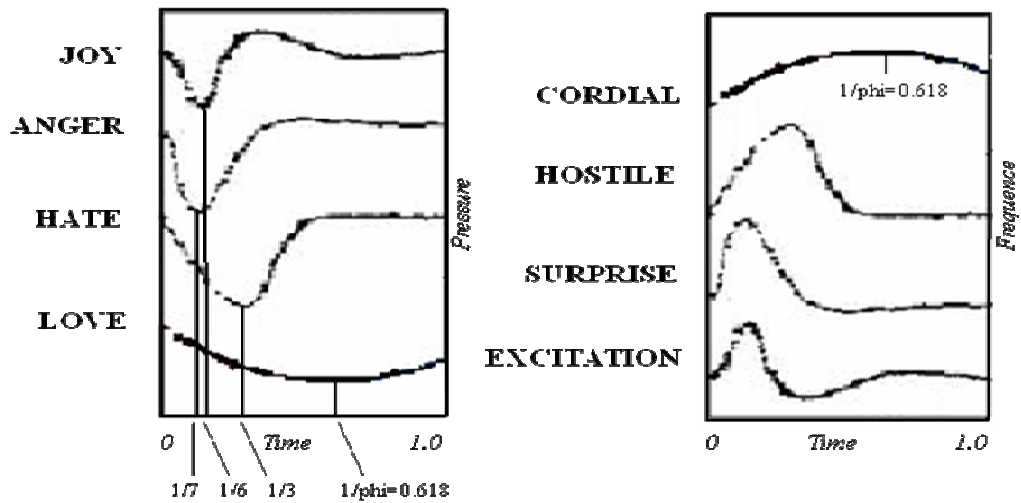

**Figure 4.** Left, the interpretation by D. Winter on some sentic forms (that correspond to four fundamental emotions). Right, the modification made by P.C. Marijuán and J.M. Buj (2007) for application to laughter, which implies an inversion of the curves and a reconceptualization of the emotions involved. The maximum corresponding to the *golden mean* may appear either to the left or to the right of the center of the curve (1/phi, 1-1/phi), depending on the "economy" of the ongoing expiration process. It has to be noted that the area subtended by the different curves is decreasing regularly in the downward direction. It means that lesser amounts of brute excitation have been minimized. See main text, and **Figure 5** too.

The crucial element to apply the sentic hypothesis to laughter is that the excursions of $F_o$ along the succession of plosives are defining the emotional *tone* of the laugh, in correspondence with one or another of the different sentic forms. According to the neurodynamic interpretation of [5], the set of variables underlying the different classes of laughter would revolve around a fundamental value: the amount of incoherent excitation instantaneously minimized. That is what the area subtended under the different classes of sentic curves means. It represents the way the excitation gradient of the global entropic variable has been handled, the kind of gradual increase and of sudden decrease suffered. This very trajectory would be manifest by means of the different emotional *tones* of the $F_o$ vibrato superimposed to the plosives. The "idle" excitation redirected toward the fonatory apparatus tells by itself what kind of gradient variation occurred during the brisk outcome of the behavioral episode. Figure 5 represents sonograms of laughter where some of these sentic forms may be detected.

In the different emotional states compatible with laughter expression, the coherence of their motor manifestations would imply that facial gestures, pitch melodies, and vocalic contents of the laughs should all of them be congruent. In the extent to which emotions such as happiness, joy, hostility, timidity or surprise are producing specific laughing signatures, they should be aligned with the other expressive components, and the resulting commonality should be susceptible of experimental checking relatively easily.

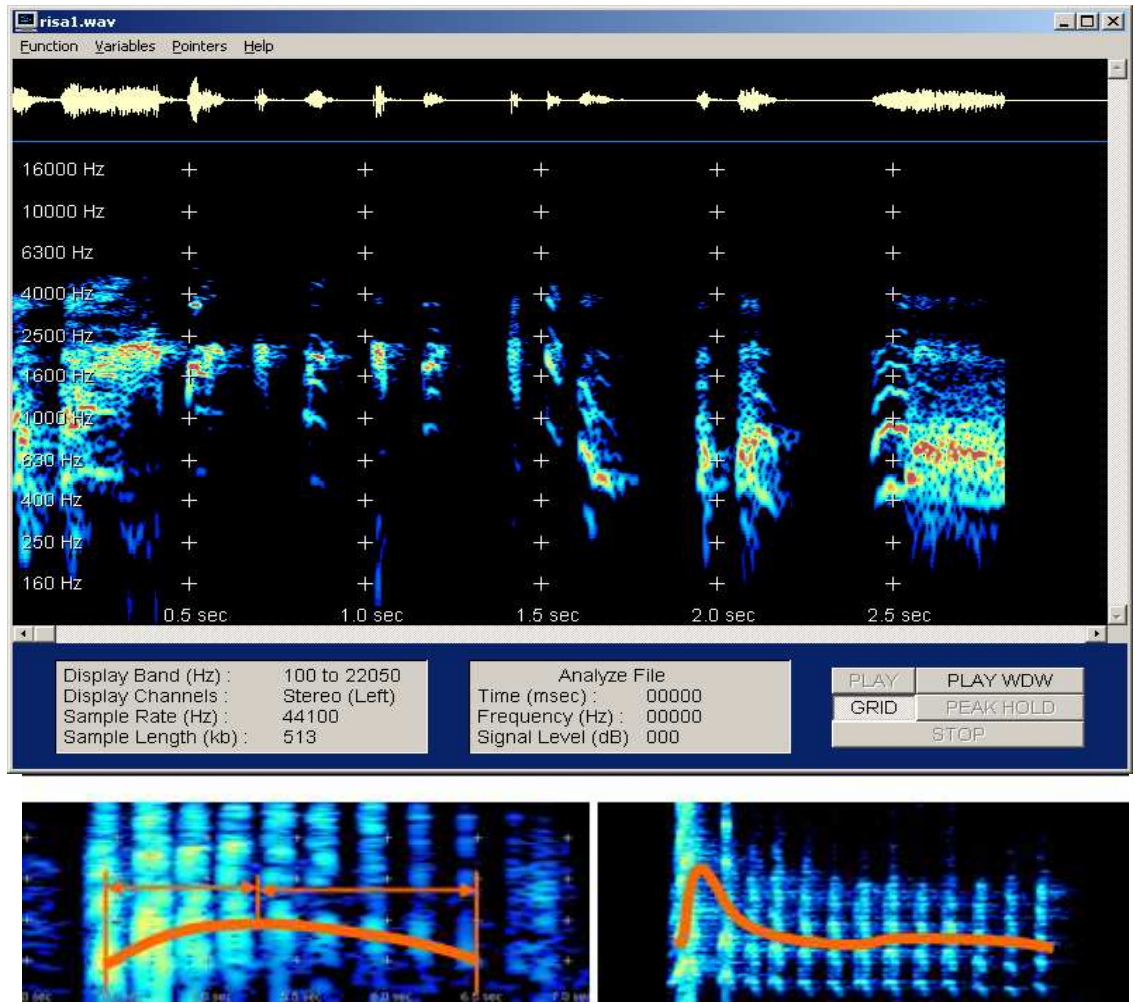

**Figure 5.** Above, sonogram of a "well-formed" laughter recorded during the joyful play of a toddler (recorded by PCM). In around 10-11 plosives it shows a crescendo, a plateau, and a decrease of the Fo values, the colors of which are graded from blue to red (from lowest to highest values); a hypothetical "arch" corresponding to the golden mean could be drawn (like in below); after this episode, a few further plosives are composing another sentic form, probably showing "excitation". Below, two sonograms are shown comprising two different sentic curves superimposed; at the left, the golden mean is appearing again; at the right, a "surprise" form is showing up, followed by a soft episode of well-formed laughter.

## 7. Conclusions: the consequences of laughter

Laughter is one of the most complex behaviors exhibited by humans. It integrates the innate and the cultural, the emotional and the cognitive, the individual and the social. Any unifying hypothesis is forced to contain an unwieldy heterogeneity of elements, even in order to attempt a very rudimentary "closure". Some of these elements may locate in well-trodden disciplinary paths and are relatively easy to discuss, while others neatly belong to the theoretical-speculative (at the time being) and become relatively disciplinary-independent. All of them, but particularly the latter, are in need of meticulous experimental approaches.

Let us summarize the main arguments herein proposed:

Human laughter, derived from primate antecedents, becomes heavily "corticalized" and associated to language, fully incorporating in this new form of social grooming as the *social brain* hypothesis has described. Laughter participates on the neuromolecular recompenses of the linguistic virtual grooming, but "augmenting" them, as it now comprises a heavy physical massage (absent in languaging) and a new form of cognitive reward throughout its "automatic" problem-solving minimization. The behavioral consequence of both the real massage and the extra endorphin reward is that the laugh signal becomes eagerly looked upon in social interactions –mainly in those where some bonding or positive memory outcome is desirable. The bonds of laughter, probably more robust as more laughter episodes accumulate upon them, will accompany the individual all along his/her life cycle: babies & toddlers, children play, adolescent groups, courtship, parenthood, grandparents, social coalitions, small-talk partners, social sharing of food...

An intriguing consequence associated to the bonding function of laughter is the conveyance of individual "identity". That's what the bonding is about: a shared cortical memory about positive interactions between *specific* individuals. In the noisy environment of the talkative human groups, the cracking sound of a highly differentiated laugh may be far more recognizable at a distance than any voiced exclamations of the same individual. Besides, it is a social signal of wellness, of bonds in the making –and exhibiting a very conspicuous signature can be interesting and advantageous in group contexts of cooperation/competition and in different stages of the individual's life cycle (e.g. specificity of materno-filial attachments). Thus, in the extent to which laugh would contain emotional signs, as well as individual cues to easily identify each subject (resembling Clynes' "personal pulse"?), a tempting speculation is that all of this could be done by tuning up on parameters of chaotic attractors in phase space.

Another promising research direction about social consequences of laughter concerns its potential use as an indicator of well-being and mental health (Hasan & Hasan, 2009), and as a diagnostic tool in neuropsychiatric pathologies, when the "bonding" capability of the individual is close to collapse (Marijuán & del Moral, 2008). An *ad hoc* research proposal has been elaborated by the authors (Marijuán, 2009).

The neurodynamic explanation of laughter herein proposed is reminiscent of R.B. Zajonc's approach to the role of CBF (cerebral blood flow) in emotional processes. In his magisterial review of Waynbaum's works on emotional expression, Zajonc (1985) argues about the pervasive role of CBF and the vascular system in mental/emotional phenomena. For instance, in social situations that cause blushing "the mobilized energy has no outlet and, as in suppressed rage, facial blood flow takes up [the discharge of] the surplus; Thus, blushing relieves CB. The face blushes... not because it is [socially] exposed but because the facial artery is a branch of the external carotid. Being constant and universal, these physiological phenomena can readily acquire communicative and symbolic significance." (Zajonc, 1985, p. 20). *Mutatis mutandis*, we are proposing a very similar approach to the neurodynamics of laughter with the redirection of the suddenly demobilized "mental energy" (brute cortical excitations), channeled toward the fonatory apparatus and toward the violent movements of intense laughing (diaphragm, respiratory, circulatory, etc.)

Of course, that this neurodynamic scheme becomes acceptable as a heuristic device is a highly debatable matter, even more in connection with the sentic forms hypothesis. But the commonality between these two views is remarkable: the global/local entropic variable comprising the evolution of brute excitation, which is shared by the different motor expression capabilities and easily recognizable by all sensory modalities. Clynes himself wrote about laughter as "another sentic form" (Clynes, 1979), or as a *composite* of sentic forms --as we would mean here. Beyond the particulars of laughter, a number of illustrious voices in contemporary neuroscience could be enlisted in support of the need of new synthetic theories about human information processing, perhaps not too distant from these argumentary lines.

Maybe another of the consequences of laughter, of its strategic placement right in the middle of human emotional-cognitive-social processes, as a safety valve of sorts, is that it shall force us to discuss on the contemporary absence of a central neurodynamic theory, about the workings of the whole cerebral cauldron.

THE BONDS OF LAUGHTER

THE BONDS OF LAUGHTER

Zajonc R.B. 1985. Emotion and Facial Efference: A Theory reclaimed. Science, 228, 15-21.